\documentclass{article}

\usepackage{arxiv}

\usepackage[utf8]{inputenc} 
\usepackage[T1]{fontenc}    
\usepackage{hyperref}       
\usepackage{url}            
\usepackage{booktabs}       
\usepackage{amsfonts}       
\usepackage{nicefrac}       
\usepackage{microtype}      
\usepackage{lipsum}
\usepackage{graphicx}
\graphicspath{ {./images/} }
\usepackage{amsmath}
\usepackage{amssymb}

\usepackage{float}

\title{goal assembly as a formalism of evolvable design}

\author{
  Dániel Czégel \\
  Y Mathematical Technologies \\
  \texttt{dc@ymathematicaltechnologies.com}}

\begin{document}
\maketitle
\begin{abstract}

In the age of artificial intelligence and biotechnology, a unified understanding of technology and biology is critically needed but still lacking. A cornerstone of such unification is evolvable design. I present a formalism, called goal assembly, that unifies three key characteristics of evolvable, biology-like design: (i) hierarchical integration of small competencies into larger competencies; (ii) highly non-uniform (power-law or log-normal distributed) representation of phenotypes by genotypes; and (iii) evolution of hierarchical modularity. It does so by making the hierarchy of physical goal states corresponding to substructures explicit, focusing on their composition across scales. In particular, higher-level goal variables approximate achievable joint goal states of interacting lower-level goal variables. Mechanisms of evolvability include goal state gradient backpropagation across scales, hierarchical decision making among alternative goal states, and structural recombination of modules. In learning theory terms, goal assembly proposes architectural inductive biases of emergent engineering systems operating at a complexity regime where goals arise as necessary abstractions. Beyond asking what goals are and how they are achieved, goal assembly suggests we can almost independently talk about their compositionality.

\end{abstract}

\newpage


\section{Introduction}

Consider what might be called the \emph{high-dimensional spherical cow} model of phenotypic evolution, where variation is modeled as a high-dimensional random unit vector. In in $d>>1$ dimensions, two random vectors are almost orthogonal; the distribution of the component of the random variation vector that's parallel to the fitness gradient limits a normal distribution with standard deviation $\sigma\sim 1/\sqrt{d}$. Then, upper bound the speed of directional evolution by selecting the single individual with the largest projection to the fitness gradient in each generation from a population size $n$. From extreme value statistics, we figure that this maximal projection scales as $\sigma\sqrt{\ln n}\sim\sqrt{\ln n}/\sqrt{d}$. This is the speed of evolution we can maximally get in the direction of the fitness gradient without any mechanism that facilitates variation to fall onto a lower-dimensional manifold. On the other hand, the projection to the direction that is orthogonal to the fitness gradient (i.e., neutral) is almost $1$. Of course, directional selection adds up over generations, and the neutral component is diffusive. Mathematically, with the number of generations $t$, evolution proceeds directionally (to the direction of the fitness gradient) as $t \sqrt{\ln n}/\sqrt{d}$, whereas in the neutral dimension as $\sqrt{t}$. It takes in the order of $d/\ln n$ generations for directional selection to take over. Mechanisms that meaningfully reduce the dimension of the search space help exponentially more than increasing population size. Arguments about high-dimensional phenotypic evolution, similar to this one, have a long history, starting with Fisher's geometric model of adaptation \cite{fisher1930genetical,orr2000adaptation,martin2006general,tenaillon2007quantifying}.

How should such dimension reduction of phenotypic variation take place? Beyond varying the length of all four legs of a chair together, there is another, cognitive, principle. Consider the not so unreal example of an adversarial agent imposing complex perturbations on your LEGO or IKEA assembly process. Soon you end up with a structure that’s \emph{not} in the manual. Luckily, you find another manual that doesn’t encode a single construction trajectory, but \emph{all} of them. It is very thick, and you end up spending hours finding a match to your structure, continuing your assembly process on that specific trajectory. Until, in five seconds, your adversary lands you on yet another new structure. The only reasonable strategy involves a paradigm shift in the creation of instruction manuals: encoding intermediate target structures, potentially multiple alternative ones that reflect the many possible parallel trajectories, and trusting the user’s common sense to reach them. While current technology does not necessarily operate at a level of complexity and perturbations where this paradigm shift is necessary, this might not be the case in the future, especially if we actively embrace such a biologyfication of technology, in this abstract, substrate-independent sense.

Illustrated by these observations, this paper aims to develop a coherent mathematical formalism for emergent engineering systems that exhibit what I call radical openness of implementational solutions across scales. Evolvability of such designs must rely on the effective integration of local, 'small' competencies across a sufficient hierarchical depth to effectively reduce design space. Generalization is crucial: in complexity and perturbation regimes where no two solutions, such as developmental trajectories, are identical, global solutions can only be reliably encoded if components can generalize across the hierarchy. This is another way of saying that components must possess competencies that abstract their possible target or goal states, which they \emph{somehow}, probably, approximately, find.

Consequently, I suggest that a minimal formalism should be built of elementary constructs, called goal variables, that encode multiple achievable goal states each, and focus on how such goal variables can be integrated across the competency hierarchy. This integration means that higher-level variables are not physical; they encode goals that are derived from local goal variables according to goal integration rules. Goal assembly aims to make this emergent hierarchy of goals explicit, encoding the global phenotype by goal composition rules (how higher level goal variables emerge when lower-level variables are composed) and their recursive compositional structure (what goal variables are composed and in what order). Figures 1a,b,c illustrate the idea.

Levin and others extensively discuss multiscale competencies in biology, including striking examples from animal development and regeneration, possible underlying microscopic mechanisms, and a cognitive modelling framework that makes emergent generalization abilities (competencies) central to understanding collective behavior of cells at various scales \cite{shreesha2023cellular, hartl2024evolutionary,pezzulo2015re,lobo2014linear,pio2023scaling, federici2006evolution, friston2015knowing, manicka2022minimal, fields2018multiscale,deshpande2024engineering,furusawa1998emergence}. This current theory I present here is strongly influenced by that line of work; what I try to do here is to formalize what competency \emph{integration} could possibly look like in physically constructed systems, targeting a level of generality that spans biological and potential emergent technological systems yet is still useful to both. If successful, one outcome of such theory building could be a more biology-like design of (everyday) objects, in the deepest sense of the word 'biological'.

I build this framework on (at least) five further domains: (i) convergent statistical properties of various genotype-phenotype maps, including models of polymer folding, self-assembly, and even parameter to function mappings in artificial neural networks \cite{schuster1994sequences, ahnert2017structural,manrubia2021genotypes,johnston2022symmetry,dingle2022phenotype,mingard2025deep}. (ii) Evolution, evolvability, and complexification of hierarchical modular structures in biology and engineering \cite{wagner1996perspective,lipson2007principles, mengistu2016evolutionary, caetano2019emergence, sharma2023assembly}. (iii) Autoassociative memories, storing a library content-addressable patterns as dynamical attractors, which proved to be a rich and effective framework for \emph{both} AI and neuroscience, and, increasingly, models of development \cite{hopfield1982neural,krotov2021hierarchical,krotov2023new,chandra2025episodic,vohradsky2001neural,watson2014evolution,paczko2024neural}. (iv) Hierarchical Bayesian models, also as a component of closed perception-action loop (active inference) models such as the free energy principle, provide an example of minimal "integration theories" (in that case, of probabilistic information) that has already proven to be successful across domains due to its appropriate level of generality  \cite{bishop2006pattern,kording2004bayesian,tenenbaum2011grow,friston2010free,friston2025active}. (v) Finally, from the side of effective dimension reduction by explicit hierarchical coarse-graining, tensor networks are perhaps the most similar in spirit, including the emphasis on structure and its graphical representation \cite{evenbly2015tensor,orus2019tensor,levine2019quantum,glasser2020probabilistic}.

\begin{figure}[H]
\centering
\includegraphics[width=1.0\textwidth]{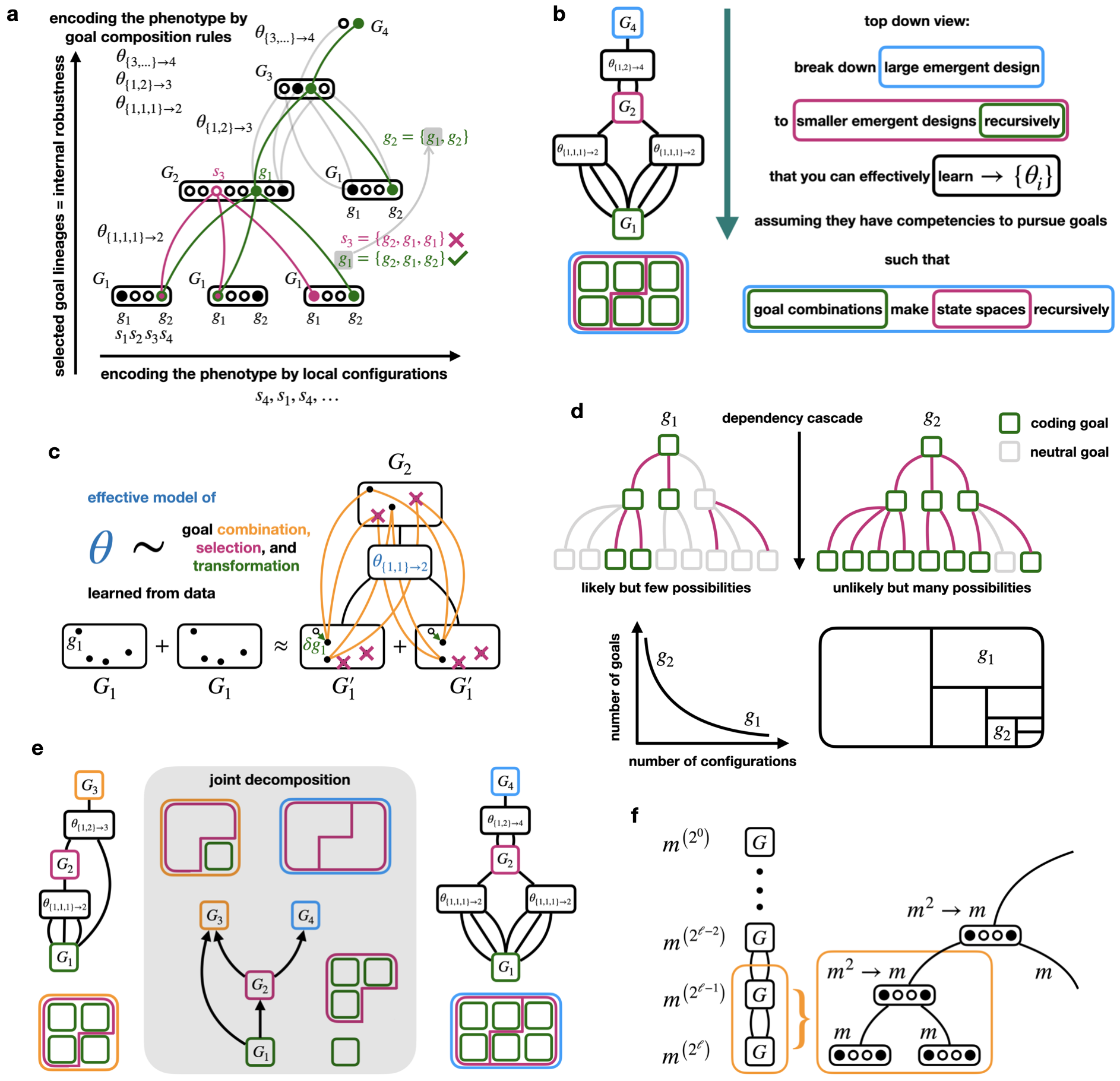}
\caption{\textbf{Goal assemblies.} \textbf{a.} Encoding the phenotype by recursive goal composition rules ("vertical view" along the hierarchy) versus local configurations ("horizontal view"). Goal composition rules $\theta$ are learned from data. Each goal variable $G$ has multiple goal states; hierarchical selection of goal lineages account for generalization-based robustness and evolvability. \textbf{b.} Top-down, or inductive, view elucidates the main assumptions behind goal assemblies: large emergent designs can be effectively decomposed to smaller emergent designs recursively, such that the resulting building blocks possess competencies to achieve their goal states. Such hierarchical reduction of model complexity is key to effective configuration (phenotype) space search. \textbf{c.} Elementary goal composition amounts to a lossy compression of the joint behavior of the physical composition of two $G_1$ goal variables. Importantly, the joint goal variable $G_2$ belongs to the same mathematical class as $G_1$, amenable for further recursive composition along the hierarchy. \textbf{d.} A generic feature of goal hierarchies is the highly uneven (power-law) representation of global goals by local configurations. This is due to a top-down dependency cascade: a single neutral goal variable makes its full corresponding subtree neutral, all the way to the leaf variables that encode local configurations. \textbf{e.} When a population of phenotypes are deconstructed, shared component goals significantly reduce joint model complexity. Such joint decomposition can serve as a basis of population-based architecture search. \textbf{f.} Example: superexponential configuration space reduction along a binary tree when each goal variable encode a sublinear number of goal states within their configuration space (here: $m^2\to m$, leading to a configuration space reduction of $m^{(2^{\ell})}\to m$ in $\ell$ steps).}\label{structure_combined}
\end{figure}

\section{Goal assembly formalism}

From a top-down perspective (Fig 1b), the core hypothesis is approximate decomposability of global design to smaller and smaller emergent competencies, interfacing each other \emph{only} through expressed (or selected) and latent goal states (Figure 1a). The emergent competencies themselves are formalized by goal variables $G$ containing goal states ${g_1,\dots g_m}$, and their interface is formalized by goal composition rules, summarized by Fig 1c. Most importantly, when goal variables are composed, lets say two $G_1$s, the \emph{state space} of the composed variable $G_2$ is composed of all combinations (the direct product) of all \emph{goal states} of the constituent goal variables $G_1$s. $G_2$ thus can be seen as a lossy compression, or approximation, of the joint behavior of two interacting $G_1$s. This interaction might cause individual goal states to be transformed (by $\delta g_1, \dots$, on Fig. 1c), and potentially selected to be also goal states of $G_2$, This makes $G_2$ amenable to be composed with another (higher-level) goal variable, all the way up the hierarchy to the top, which represents the global target structure (phenotype).

For a specific physical realization, say in the previous example of $G_1+G_1\approx G_2$ shown on Fig. 1c., all this information might not be deduced but learned (approximated) from data, obtained from physical experimental realizations or more detailed simulations. This mapping is denoted by $\theta_{1,1\to 2}$. To avoid confusion, I emphasize again that this is still not a model of how goals are achieved, but only a parametrized model of the  composite behavior $G_2$ given two $G_1$s are combined (in a specific way). Practically, this might lead to a library of effective models of goal composition rules $\theta_{1,1\to 2}, \theta_{1,2\to 3}, \dots$, which then can be used to \emph{guess} the behavior of novel (artificial or evolved) combinations of substructures, as illustrated by Fig. 2c.

An emerging universal feature of emergent engineering systems, and models of them, such as genotype-phenotype maps and multilayer neural networks is what has been called an “automatic Occam’s razor”: phenotypes that have a simple effective computational description are represented by a larger fraction of genotype (or description, or blueprint, or input) space, than those with a complex effective description \cite{johnston2022symmetry, dingle2022phenotype, mingard2025deep}. It is important to emphasize that these are effective computational models that are not necessarily algorithmically realized or even approximated by the actual unfolding of the phenotype from the genotype. This representational bias is quantified by the distribution of the number of genotypes that map to a phenotype, with a few phenotypes (with a simple effective computational description) being represented by a large fraction of the genotype space, and many (complex) phenotypes are each represented by a small fraction of it. This representational bias distribution takes the form of a power law or a log-normal distribution, observed across a wide variety of systems \cite{schuster1994sequences, ahnert2017structural,manrubia2021genotypes,johnston2022symmetry,dingle2022phenotype,mingard2025deep}. A recent study \cite{mingard2025deep} of a paradigmatic emergent engineering system outside of the realm of model genotype-phenotype maps nicely illustrates the point. When multilayer neural networks are initialized with random weights (i.i.d. from a normal distribution with a large enough variance), although different realizations might execute the same series of \emph{actual} computational steps on each input to arrive at the output, their effective algorithmic complexity (approximated by compression algorithms such as Lempel-Ziv \cite{ziv1977universal}) strongly varies, with algorithmically simpler input-output maps being represented by a large number of realizations, whereas complex effective maps are represented by a much smaller number of them.

There might be multiple non-conflicting explanations of this phenomena; I would like to point out how it might be, at least qualitatively, explained by \emph{dependency cascades} in hierarchical systems. First, representational degeneracy, i.e., the number of local configurations that encode a given global configuration, is bounded by the number of \emph{non-coding} local variables. For example, if a global configuration (phenotype), built from $N$ local variables overall, is encoded by only $C$ of the $N$ local variables, the degeneracy is at least $\prod_{i=1}^{N-C} m_i$ where $m_i$ are the number of possible states each non-coding local variable $i$, which is exponential in $N-C$. One way to have phenotypes with strongly varying degeneracies is to have a large enough variation in the number of non-coding local variables $N-C$. This can be achieved in hierarchical systems by dependency cascades: when a given goal variable is independent of at least one of the constituent goal variables, it is automatically independent of the full subtree of descendant goal variables, all the way down to the elementary goal variables. This has an exponential amplification effect: the higher in the hierarchy this independency happens, the larger the downstream effect is, resulting in more non-coding local variables. Writing the typical number of goal states per variable $m$ is as $m=e^\alpha$, the size of local configuration space $s$ that represents a phenotype with $C$ coding local variables and $N-C$ non-coding ones is $s=e^{\alpha (N-C)}$, or equivalently, $C=N-\alpha^{-1} \ln s$. Then, from the transformation rule of probability density functions, $P(s)=P(C(s))|dC/ds|$, with $|dC/ds|=(\alpha s)^{-1}$. This, non-trivially, adds a "Zipfian" baseline to the representational bias distribution. For example, when $P(C)$ is uniform, $P(s)\sim s^{-1}$. Note that uniform $P(C)$ can be easily realized with a (near-)optimal encoding scheme, like prefix codes assigned by Huffman coding; however, since this is an optimization process involving the whole genotype space, it cannot be selected for, it must be emergent. What I offer here is a possible mechanistic explanation, using dependency cascades across the hierarchy, generating a wide variety of local coding variable numbers $C$ across phenotype realizations. This variety is, in turn, amplified when transformed to the share of genotype space $s$. Consider a simple and conservative toy model, where the number of coding local variables $C$ is Poisson distributed, $P(C) \propto \frac{\lambda^C}{C!}$, with a standard deviation that scales as the square root of the mean $\lambda$. This can also be thought of as a lower bound on the width, generated by a \emph{non-cascading} process, where the global phenotype depends on every local variable with a constant small probability. Any cascading process likely increases the width of this distribution. Since there is no reason to assume any further structure to the dependency cascade at this point, I provide an exact calculation of this lower bound here. Transforming $P(C)$ to $P(s)$ gives $P(s) \propto \frac{s^{-\alpha^{-1} \ln \lambda - 1}}{(N-\alpha^{-1} \ln s)!}$ (see Methods) which is a power law with a factorial cutoff, well approximated by the power law $P(s)\approx s^{-\alpha^{-1} \ln \lambda - 1}$ when $s \lesssim e^{\alpha N}$, which is, approximately, the size of the whole genotype space. In other words, this "Zipfian baseline" seems to dominate the qualitative behavior of $P(s)$, but does not give us the numerical value of the exponent.

Intuitively, compared to the configuration space, the "number of local variables" space is logarithmic, and the depth of the hierarchy is logarithmic in the number of local variables, and is double logarithmic in the number of configurations. This has a generic super-exponential or double-exponential configuration space reducing effect in the depth of the hierarchy. One simple example of that is shown by Fig 1f, where each goal variable selects $m$ of $m^2$ states as goal states, arranged in a binary tree, leading to a configuration space reduction of $m^{(2^\ell)}$ in $\ell$ hierarchical integration (or goal composition) steps.

Another core feature of goal assemblies is assumed to be a specific type of hierarchical modularity, in which components are often identical, as shown by Fig. 1e. "Vertical" encoding, using goal composition rules, as opposed to local configurations (Fig. 1a.), naturally accounts for this redundancy. This allows evolutionary innovation to happen in the space of compressed representations given by the goal integration rules. This kind of reliable encoding based innovation dynamics in the (open-ended) space of hierarchical modular structures is proposed to be a main focus of assembly theory \cite{sharma2023assembly}, which is an additional reason why I call this current formalism goal assemblies, beyond the colloquial use of the word 'assembly'. Evolution in compressed representation space unites this 'pillar' of goal assemblies with the previous one concerning statistics of genotype-to-phenotype or blueprint-to-construct maps (Fig. 1d.). Furthermore, joint decomposition of multiple phenotypes, as illustrated by Fig. 1e., can serve as a coordinate system, or map, which can be then used to formulate models of innovation dynamics in this compressed representation space. Historically contingent complexification dynamics ("in order to build something complex that works, you first need to build something simple that works"), with re-use and recombination of modules, here modules being minimal competencies, is a natural place where this framework might prove to be useful. As Fig. 2d. illustrates, evolvability in architecture space relies on a formalism that maps the 'adjacent possible' of a population of hierarchical modular competency structures appropriately; goal assemblies provide a principled, compressed representation based approach.

In this sense, goal assembly unifies three aspects of biological organization: (i) hierarchical integration of (small, emergent) competencies, (ii) representational bias in genotype-phenotype maps, (iii) joint hierarchical modularity. Let's now focus on how goal assemblies evolve, emphasizing mechanisms of generalization-based evolvability at various scales.

\begin{figure}[H]
\centering
\includegraphics[width=0.82\textwidth]{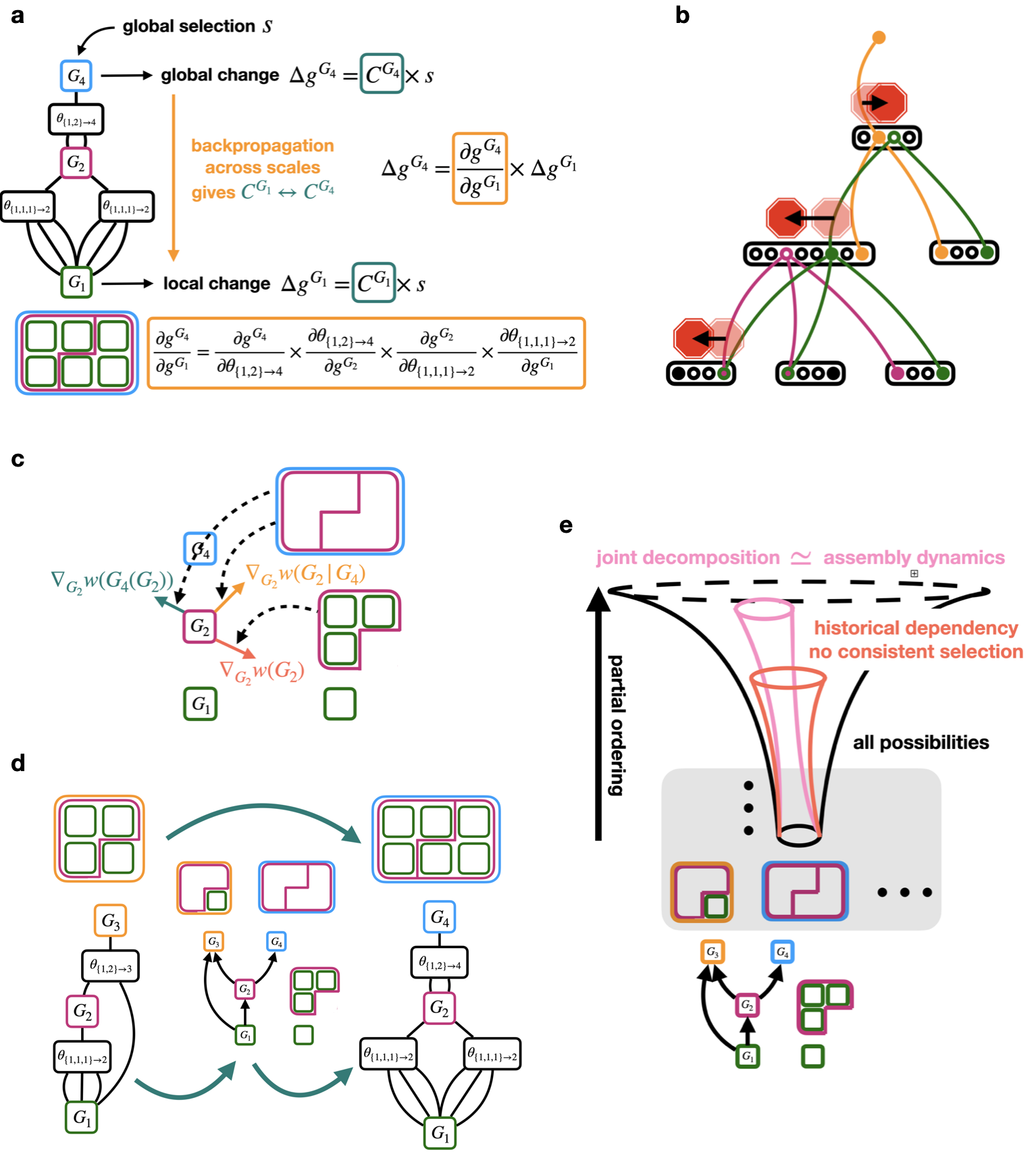}
\caption{\textbf{Evolvability of goal assemblies.} \textbf{a.} Backpropagation of phenotypic change across scales, from global ($\Delta g^{G_4}$) to local ($\Delta g^{G_1}$), through the computational graph formed by chains of goal composition rules.  Alternatively, the phenotypic trait covariance matrix $C$ can also be backpropagated, describing small continuous response to selection. \textbf{b.} Hierarchical decision making among existing goal states as a mechanism of evolvability. Goal states represent a small subset of possible configurations that have been a direct or indirect consequence of past selection, akin to latent (hierarchical) memories. The global phenotype changes when local goal state changes propagate all the way up the hierarchy. When such novel global goal states are constructed, instead of directly selected for, this mechanism amounts to a second type of competency, namely goal induction (the first being reaching the goal states themselves under novel perturbations). \textbf{c.} When goal variables are given evolutionary autonomy, conflict between levels can arise: selection at the “collective” level $G_4$ induces an evolutionary change on $G_2$, $\nabla_{G_2} w(G_4(G_2))$, other than $G_2$’s self interest \emph{within} $G_4$, $\nabla_{G_2} w(G_2|G_4)$, and yet another when $G_2$ are free, $\nabla_{G_2} w(G_2)$. (For convention and simplicity, evolutionary response to selection is represented on this panel by a gradient of the local "fitness landscape" in question, $\nabla w$.) When designed, evolutionary autonomy of goal composition rules (or goal variables), might lead to the evolution of top-down regulation as a response to the conflict of interest between levels as well as a potential emergence of higher-level evolutionary individuality, coinciding with a higher-level unit of competency in this precise sense of goal variables, all “from within”. \textbf{d.} Structural evolution of goal assemblies. Joint decomposition of a population of phenotypes and their possible single-step structural neighborhood maps the space of possibilities, which can be then used to train parametrized models to enhance evolvability. Here, a structural intermediate is shown between a current phenotype its one possible structural mutation. \textbf{e.} Accounting for recursively shared parts makes it possible to scale up structural generalization (as in d.) to a population of complexifying architectures. Statistical rule classes of assembly can be used both as a generative process and as descriptive statistics inferred from the joint decomposition of existing (observed) architectures.}\label{evolution_combined}
\end{figure}

\section{Evolvability of goal assemblies}

In the following, we discuss three mechanisms of evolvability, beyond goals themselves being the “fundamental unit” of generalization. One is directly backpropagating selection gradient \cite{rumelhart1986learning, bengio2017deep} through the computational graph of goal composition rules, as shown by Fig 2a.  Alternatively, one can imagine phenotypic trait covariance matrix to be backpropagated. Indeed, goal variables and their composition rules gradually transform local configurations to the global phenotype, and this is true in a continuous, variational sense, too: selecting on one phenotypic dimension indirectly selects on the others, which is exactly what the phenotype trait covarince matrix captures. Second, hierarchically selecting between alternative goal states can reduce search space effectively without structural recombination or inventing new modules. This is akin to accessing latent memory states \cite{hopfield1982neural,krotov2021hierarchical}, although here we are agnostic about how they have been acquired in the first place.  Figure 2b illustrates the idea. Third, due to the explicit description of joint hierarchical modularity, structural recombination and long timescale complexification of goal hierarchies can be inferred or modelled, as shown by Figures 2d and 2e. Indirect encoding of the phenotype by hierarchical goal composition rules is also consistent with bottom-up evolution of individuality, whereby multilevel selection mechanisms, as illustrated by Figure 2c, might induce a transition in individuality, being a major engine of complexification \cite{smith1997major, okasha2005multilevel, west2015major}.

Goal state gradient backpropagation across scales. In representations where goal states and their composition rules are effectively differentiable, goal state gradient backpropagation might provide a mechanism of evolvability through computational speedup of credit assignment, similar to the computational speedup in training of multilayer neural network architectures. From an evolutionary perspective, computationally efficient reverse engineering could be considered a \emph{maximally Lamarckian} mechanism, encoding desired phenotypic change in the changes in the encoding of the parts (here, in goal composition rules, parametrized by $\theta$s). In goal assemblies, as opposed to neural networks, backpropagation operates across scales, from global to local, as Fig 2a illustrates. Regardless of how global phenotype dimensions, or traits, are defined in any specific case (for example, based on practicality and measurability), they somehow map to local goal state configurations; tracking the way goals transform across the hierarchy amounts to tracking transformations of phenotypic traits. Specifically, when the effect of global selection $s$ is backpropagated across goal states, it is also backpropagated across the phenotypic trait covariance matrix $C$ that describes the effect of selection on phenotypic dimension $i$ on phenotypic dimension $j$.

Figure 2b illustrates the second mechanism of evolvability, hierarchical decision making among goal states. This can be regarded as an analogue of hierarchical associative memory retrieval: existing goal states as direct or indirect consequences of selection history, are generalizations of learned memory traces. In cognition, encoding and retrieval of hierarchical memories form a key mechanism connecting past experience to (drivers of) present behavior. When some of the alternative goal states have not been directly selected for, this mechanism amounts to what we might call \emph{goal induction}, an “educated guess” of alternative viable phenotypes. In this precise learning theory sense, they are generators of novelty. Since phenotypes are encoded as a series of goal composition rules that instruct their assembly, goal induction necessarily operates under a compositional inductive bias. Mechanistically, phenotypes change when changes at any level (scale) of the hierarchy propagate all the way up to the global goal variable. Otherwise, they are neutral; in this sense, neutral spaces are large enough to host a meaningful cryptic (hidden) variation, further biasing phenotypic mutation probabilities.

Multilevel selection \& transition in individuality.
Goal variables might be units of selection themselves, with various degree of evolutionary autonomy. This is similar in spirit to previous formal attempts to relate agential and evolutionary units of individuality, see, for example, \cite{krakauer2020information,watson2022design,tissot2024ability}.
In goal assemblies, this is made explicit by the “vertical”, goal composition rule based encoding of the phenotype. As shown by Fig 2c, at least three distinct evolutionary forces can act upon a generic goal variable, $G_2$. When these selective forces are represented by fitness gradients $\nabla w$ for simplicity, these are $\nabla_{G_2} w(G_2)$ when $G_2$s are evolutionarily autonomous (free), $\nabla_{G_2} w(G_2|G_4)$, the selective force on $G_2$ \emph{within} $G_4$, and $\nabla_{G_2} w(G_4(G_2))$, the selective force on $G_2$ as (passively, in an evolutionary autonomy sense) as part of $G_4$. The difference between the last two makes it possible to represent partial autonomy, which is a stepping stone (or, rather, a continuous ramp) towards a transition of evolutionary individuality. Successive transitions in individuality might be one mechanism of emergent (not externally designed, “from within”) complexification of goal assemblies.

More generally, regardless of the evolutionary force, structural evolution of goal assemblies can be tracked by a graphical (and mathematical) representation, such as the one shown on Fig 2d. When a population of  phenotypes is considered, their (approximate) joint decomposition into shared components, and shared components of components, allows for mapping the space of possible recombinants \cite{liu2021exploring,sharma2023assembly,patarroyo2023assemblyca}. This accounts for yet another type of induction-based evolvability, where existing modules are recombined in a principled way. Compositional (“vertical”) encoding serves as a natural representation. Outer-loop models that assign probabilities to structural mutations might be able to meaningfully prune recombination space further. I suggest structural intermediates, as shown on Fig 2d, as representations of appropriate complexity that can serve as a basis of training such architecture search models.

When structural evolution is extrapolated to longer timescales, different classes of complexification dynamics can arise (Figure 2e). This can be mapped by (i) partial ordering recursively shared parts, and (ii) using the so-created “ladder”, or co-ordinate system, to map evolutionary trajectories within the space of all possibilities. As suggested in \cite{sharma2023assembly}, approximating the complex map of parts by simple models of forward dynamics ("joint decomposition $\approx$ assembly dynamics" on Figure 2e) classifies possible complexification trajectories. In particular, consistent selection can nudge the system to explore an even smaller (or larger) fraction of a complexity "slice" than it would do with historical contingency alone, depending on what is desired in the emergent engineering system at hand. I view this as a longer timescale version of structural evolvability discussed in Fig 2d. Relating local bias in recombination space to global bias in assembly space in a principled manner would be an important bridge between these two evolutionary scales.

\section{Discussion}

Goal assemblies might be considered to be part of a broader project, which I call the 'recursification of biology'. The idea is to try to account, at the same time, for multi-scale, cumulative, and heterogeneously implemented emergence, with higher levels of organization building on selected reliable 'condensed' units of lower-level organizations. Perhaps a good analogy is the emergence of symbolicity and grammar in language, from analog undifferentiated to discrete and more and more reliable, and, ultimately, grammatical, with a flexibly controllable, practically infinite combinatorial possibility space. This is an analogy, but illustrates the point: when this infinite possibility space and the analog space to begin with is treated in the same framework, this process can unfold recursively. Again, think of this process in the non-predefinable space of heterogeneous, physically implemented engineering solutions.
The right level of abstraction is key: there might be space for useful mathematics here, might be not; I predict (and advocate for) the former. 
From a modeling perspective, what such models prescribe is the integration of hypothesized structure with learned effective models of particulars, instead of being fully predictive on their own. In learning theory terms, they might be called 'architectural inductive biases'.  Furthermore, I highlight the importance of accounting for the richness and full breath of recursive construction versus simpler (e.g., layer-by-layer) hierarchies. Finally, such heterogeneous, cumulatively unexpected engineering solutions might be found only by direct physical experimentation and not planning via simplified models. Unifying such experimentation (evolutionary) history with structure, on the basis of historical contingencies and counterfactuals, might be a necessary component of the theory of biology-like design.

\section{Methods \& notes}

\subsection{Spherical cow}

At the limit of infinite dimensional Euclidean phenotype spaces and infinite population size, the following precise argument can be made. The 'projective central limit theorem' says that the one dimensional marginal of a uniform distribution over the unit sphere in a $d\to \infty$ dimensional Euclidean space converges to a normal distribution with standard deviation $\sigma=1/\sqrt d$, see, for example, the excellent book \cite{vershynin2018high}. Then, the \emph{mean} of the distribution of the maximum of $n\to\infty$ i.i.d. samples from a standard normal distribution scales as $\sqrt{2 \log n} - (\log \log n + \log 4\pi)/(2\sqrt{2 \log n})$ 
 \cite{coles2001introduction}, with leading term $\sqrt{2 \log n}$.

\subsection{Poisson with Zipfian baseline}

Substituting $C(s) = N-\alpha^{-1} \ln s$ and a Poission distribution $P(C) = \frac{\lambda^C e^{-\lambda}}{C!}$ to the transformation rule $P(s) = \frac{1}{\alpha s} \cdot P(C(s))$ gives

\begin{equation}
P(s) = \frac{1}{\alpha s} \cdot \frac{\lambda^{N-\alpha^{-1} \ln s} e^{-\lambda}}{(N-\alpha^{-1} \ln s)!}
\end{equation}

This can be simplified using $\lambda^{-\alpha^{-1} \ln s} = e^{-\alpha^{-1} \ln s \ln \lambda} = s^{-\alpha^{-1} \ln \lambda}$ to get

\begin{equation}
P(s) = \frac{\lambda^N e^{-\lambda}}{\alpha s \cdot (N-\alpha^{-1} \ln s)!} \cdot s^{-\alpha^{-1} \ln \lambda} \propto \frac{s^{-\alpha^{-1} \ln \lambda - 1}}{(N-\alpha^{-1} \ln s)!}
\end{equation}

This distribution has two regimes: for sufficiently small $s$, the distribution is well approximated by a power law with a different exponent than the baseline exponent $\alpha$,

\begin{equation}
P(s) \propto s^{-\alpha^{-1} \ln \lambda + 1}
\end{equation}

until a cutoff at $s = e^{\alpha N}$, beyond which the factorial term dominates.

\section*{Acknowledgements}

I thank Sara Walker, Eörs Szathmáry, Hamza Giaffar, Dániel Barabási, and Veronica Mierzejewski for extensive discussions that partly inspired this project.

\bibliographystyle{unsrt}  
\bibliography{goal_assembly}

\end{document}